\documentclass[%
reprint,
twocolumn,
superscriptaddress,
graphicx,
floatfix,
amsmath,
amssymb,
bibnotes,
aps,
prb,
]{revtex4-2}
\usepackage[]{graphicx}
\usepackage{epstopdf}
\usepackage{MnSymbol}
\usepackage{latexsym}
\usepackage{color}
\usepackage{dcolumn}
\usepackage{bm}
\usepackage[mathlines]{lineno}
\usepackage{usebib}

\begin{document}

\preprint{Submitted to Optics Letters}

\title{Temperature-dependent second-harmonic generation from color centers in diamond}

\author{Aizitiaili Abulikemu}
\affiliation{Department of Applied Physics, University of Tsukuba, 1-1-1 Tennodai, Tsukuba 305-8573, Japan.}

\author{Yuta Kainuma}
\affiliation{School of Materials Science, Japan Advanced Institute of Science and Technology, Nomi, Ishikawa 923-1292, Japan.}

\author{Toshu An}
\affiliation{School of Materials Science, Japan Advanced Institute of Science and Technology, Nomi, Ishikawa 923-1292, Japan.}

\author{Muneaki Hase}
 \email{mhase@bk.tsukuba.ac.jp}
\affiliation{Department of Applied Physics, University of Tsukuba, 1-1-1 Tennodai, Tsukuba 305-8573, Japan.}

\begin{abstract}
Under infrared ultrashort pulse laser stimulation, we investigated temperature-dependent second-harmonic generation (SHG) from nitrogen-vacancy (NV)-introduced bulk diamond. The SHG intensity decreases in the temperature range of 20--300$^{\circ}$C, due to phase mismatching caused by refractive index modification. We discovered that optical phonon scattering outperforms acoustic phonon one in NV diamond by fitting the temperature dependence of SHG intensity using a model based on the band-gap change via the deformation potential interaction. This study presents an efficient and viable way for creating diamond-based nonlinear optical temperature sensing.
\end{abstract}

\date{\today}

\maketitle
Temperature sensors are used in various electronic devices such as air conditioners, refrigerators, automobile engines, and personal computers, which require measurement accuracy and stability. There are various types of temperature sensors, which can be classified into contact and noncontact types. Contact type temperature sensors often include resistance thermometers, thermistors, and thermocouples, whereas noncontact type sensors are mainly quantum sensors \cite{Benedict}. Noncontact quantum sensors with nitrogen-vacancy (NV) centers in diamond are based on the principle that the resonant microwave frequency of the luminescence between quantum levels of the NV center varies with temperature \cite{Wang2015}, and is expected to be applied to intracellular measurements requiring high spatial resolution and high sensitivity, as well as sensors for microscopic evaluation devices.

Diamond NV center is an atom-like defect [Fig. 1(a) inset], consisting of a substitutional nitrogen atom and an adjacent lattice vacancy next to the carbon atom \cite{Pezzagna}. Ion implantation and high-temperature annealing are commonly used to introduce NV centers at the surface (several tens of nanometers deep). Recently, NV color centers in diamond have received great interest owing to their rich photophysical properties and potential applications, including light sources \cite{Breeze,Mildren}, photonic device technologies toward quantum computations \cite{OBrien,Sekiguchi,Janitz}, single-photon sources \cite{Babinec,Mizuochi,Aharonovich}, and quantum sensing \cite{Kim2019,Pelliccione}. Furthermore, quantum sensing using NV centers in diamond has attracted much attention and has been used to measure electric fields \cite{Dolde}(current) and magnetic fields \cite{Maze2008} (spin), as well as temperature sensors \cite{Kucsko,Clevenson}. In contrast, crystal symmetry, most importantly the presence of inversion symmetry, plays a critical role in determining the optical properties of materials \cite{Trojnek}. NV diamond contains an NV color center near the surface region that could break the centrosymmetric structure in diamond crystals since the N atom and its adjacent vacancy break the inversion symmetry \cite{Abulikemu,Motojima} and thus develop second-order nonlinear optical (NLO) effects—the second-harmonic generation (SHG) \cite{Abulikemu}. Although much attention has been devoted to the magnetic properties of NV diamond, there have been few reports on the NLO properties of NV diamond crystals \cite{Abulikemu,Motojima,Angerer}.

In this study, we investigated second-order NLO properties of diamond with NV color centers in terms of temperature dependence and found that SHG and SHG-induced cascaded third-harmonic generation (cTHG) are simultaneously observed when excited by an infrared (IR) ultrashort laser pulse. According to nonlinear optics theory, the second-order nonlinear polarization of SHG can be described as  $P_{SHG}$= $\chi^{(2)}$$EE$, where $\chi^{(2)}$ is known as the second-order NLO susceptibility, and $E$ is the driving electric field. The SH output intensity $I$ can be expressed as \cite{Shen}:
\begin{equation} \label{eq1}
I(2\omega) \propto |\chi^{(2)}(2\omega;\omega+\omega)|^{2}sinc^{2}
\bigl(\frac{\Delta kz}{2}\bigr)
 I(\omega)^{2},
\end{equation}
where $\omega$ and 2$\omega$ are the carrier frequency of the excitation and SH signals, respectively. $\Delta k$ and $z$ are the phase mismatch and 
propagation axis, respectively. The SHG intensity becomes maximum when the phase-matching condition of $\Delta k$=0 or $n_{2\omega}$ = $n_{\omega}$ is 
satisfied as demonstrated by Eq. (1), where $n_{2\omega}$ and $n_{\omega}$ are the refractive index for the SHG and fundamental wavelengths, respectively. 
Furthermore, carefully designed optical systems can allow generating cTHG signals by a two-step (or cascading) process driven 
by $\chi^{(2)}$$:\omega$+$\omega$ $\rightarrow 2\omega$ (SHG), and then: 2$\omega+\omega \rightarrow 3\omega$ (cTHG) under perfect phase-matching conditions \cite{Shen}.

\begin{figure}[h]
   \begin{center}
    \includegraphics[width=7.5cm]{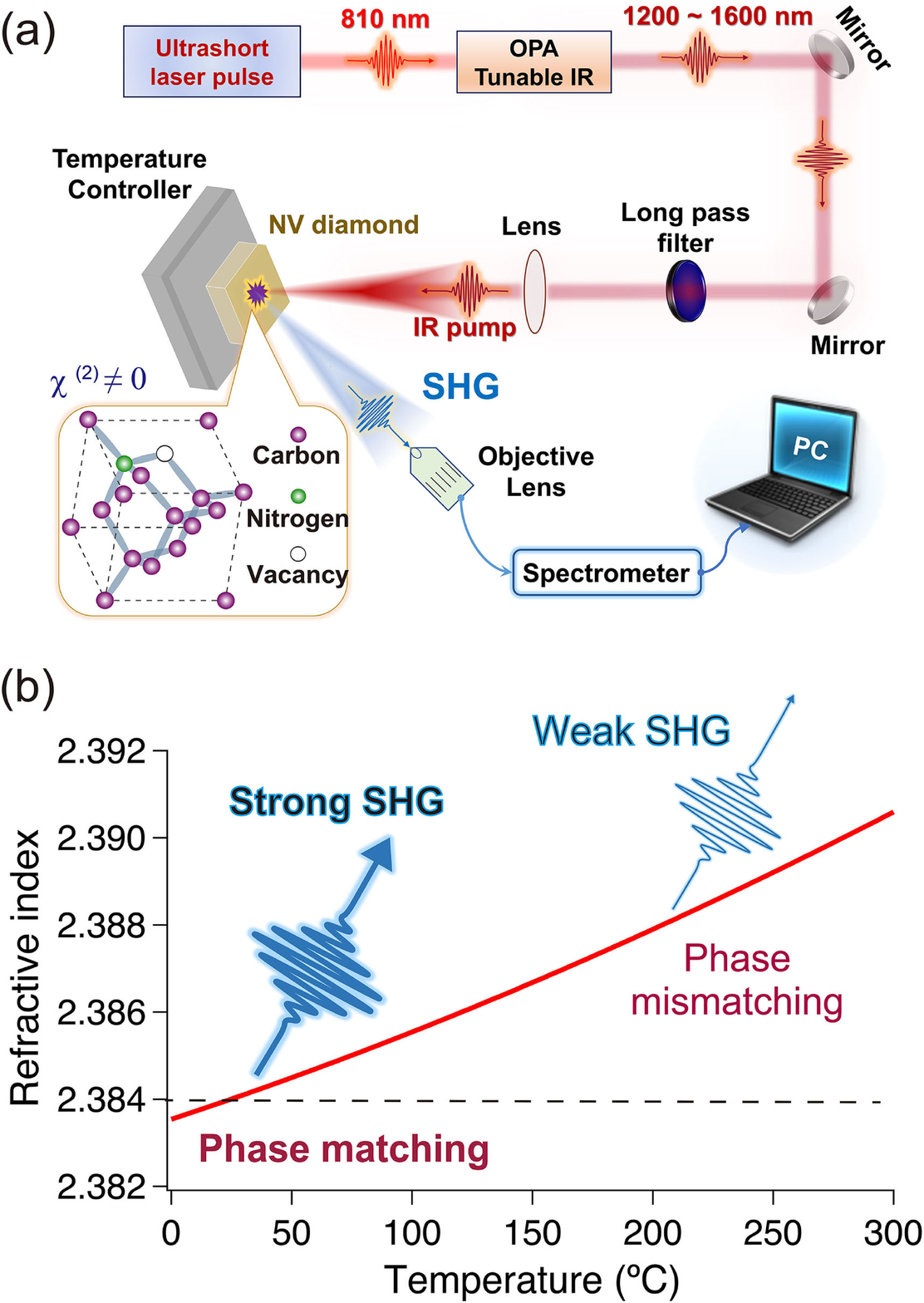}
    \caption{(a) A simplified diagram of the reflection type measurement scheme. The inset shows atomic structure of the nitrogen-vacancy (NV) center in a diamond crystal. (b) A relationship between the temperature ($T$), refractive index ($n$), and the SHG intensity in NV diamond crystals. The refractive index (red curve) is calculated using the 3rd-order polynomial function of  $n(T)=n_{0}+aT+bT^{2}+cT^{3}$, where $n_{0}$= 2.38, $a$ = 7.47$\times$10$^{-6}$, $b$ = 2.09$\times$10$^{-8}$, and $c$ = 2.92$\times$10$^{-12}$ are taken for the CVD diamond \cite{Yurov}. The horizontal dashed line represents the phase-matched refractive index at room temperature.
    }
   \label{FIG1}
   \end{center}
  \end{figure}
  
In the following, the unique temperature dependence of SH signal results from phase mismatching caused by the modification of the refractive indexes ($n_{2\omega}$ and $n_{\omega}$) by band-gap modulation via deformation potential electron-phonon interaction in NV diamond crystals. To the best of our knowledge, the temperature dependency of the SHG in NV diamond has never been established, and this is the first work attempting to apply NLO effects-based temperature sensing devices using NV diamond crystals. Moreover, we have investigated the wide-band wavelength tuning characteristics of nonlinear emissions to increase the possibility of a diamond photonic wavelength-tunable laser source. 

The sample used was an (100)-oriented single-crystal diamond grown by chemical vapor deposition (CVD) with a size of 3.0 $\times$ 3.0 $\times$ 0.3 mm (thickness). The NV diamond crystal hosting NV centers about 30--40 nm-deep was the same sample in the previous study \cite{Abulikemu} and prepared by combining the ion implantation (30 keV at a dose of 5.0 $\times$ 10$^{12}$ ions/cm$^{2}$) and high-temperature annealing in argon atmosphere \cite{Prananto}. The presence of ensembles of NV centers in diamond crystals was demonstrated by the confocal scan image of photoluminescence as well as by the measured broadband photoluminescence spectrum with the characteristic zero phonon lines \cite{Abulikemu}. 

A simplified optical system for measuring nonlinear emissions from NV diamond crystals is shown in Fig. 1(a). The NLO spectra were captured using a home-built optical setup in reflection geometry at nearly normal incidence. Femtosecond laser pulses from a mode-locked Ti:sapphire oscillator (pulse width 20 fs; central wavelength 810 nm; repetition rate 80 MHz; and average power 425 mW) were amplified by a regenerative amplifier to obtain laser pulses at 100 kHz. The output pulses were fed into an optical parametric amplifier (OPA), which generated IR pulses with wavelengths ranging from 1200 to 1600 nm and a maximum IR signal power of 100 mW. IR excitation beams were spectrally cleaned with a long-pass filter, the power of the exciting laser pulse was controlled by a neutral density filter, and the excitation beam was focused on the sample surface with a focusing lens ($f$ = 50 mm) to a beam diameter of $\sim$20 $\mu$m. During measurements, the excitation fluence of $I_{ex}$ = 7.6 mJ/cm$^{2}$ was kept constant under IR excitation centered at $\lambda_{ex}$ $\sim$ 1350 nm. The NV diamond sample was mounted on an oven, and the closed-loop temperature controller enabled us to change the crystal temperature. The emission signals were collected by a collimating objective lens to couple into an optical fiber and finally detected and integrated for 30 seconds by a spectrometer. As demonstrated in the following, both SHG (peak at $\sim$ 675 nm) and cTHG (peak at $\sim$ 450 nm) signals are simultaneously emitted from the symmetry-breaking nanometer-scale NV layer and show significant changes with increasing crystal temperature [Fig. 1(b)] \cite{Yurov}.

  \begin{figure}[h]
     \includegraphics[width=7.7cm ]{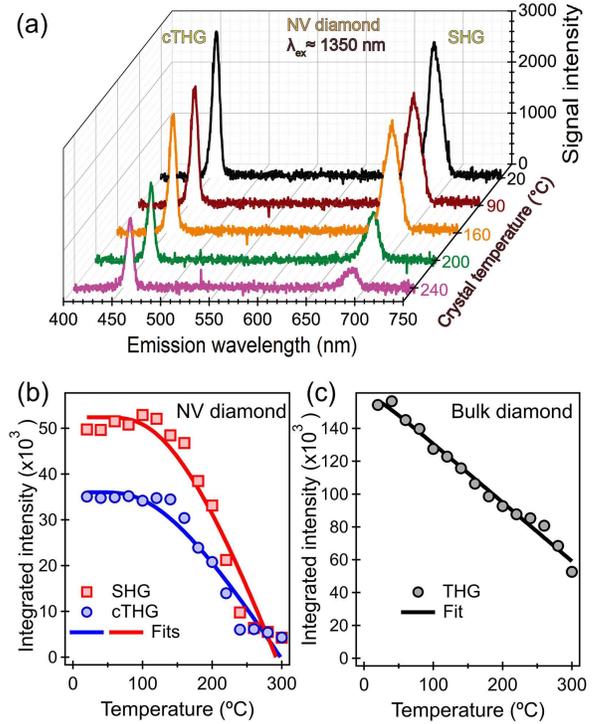}
     \caption{(a) Experimental results of second-harmonic generation (SHG) and cascaded third-harmonic generation (cTHG) spectra as a function of the crystal temperature for five values: black, deep red, orange, green, and purple lines corresponding to 20$^{\circ}$ (room temperature), 90$^{\circ}$, 160$^{\circ}$, 200$^{\circ}$, and 240$^{\circ}$ respectively. (b) The temperature tuning curve of the SHG (red squares) and cTHG (blue circles) outputs at various temperatures. (c) Temperature dependence of the THG intensity from bulk (pure) diamond crystal. The solid lines are the fits using Eq.  (2) plus a background.  The values of $\alpha$ obtained for the NV diamond are 1.22 $\times$10$^{5}$ and 5.65 $\times$10$^{4}$ for SHG and cTHG, respectively, and 1.93 $\times$10$^{4}$  for THG in bulk diamond. }
     \label{FIG2}
     \end{figure}

Figure 2(a) shows the typical nonlinear emission spectra obtained from an NV diamond at various temperatures, from 20$^{\circ}$C (room temperature) to 
240$^{\circ}$C, where nearly perfect phase matching was elaborated by adjusting the angle of the birefringent NV diamond sample at 20$^{\circ}$C \cite{Boyd}. 
In the low-temperature regions, the detected SHG and cTHG intensity were almost the same and gave the optical conversion efficiency, respectively. Upon heating of the NV diamond crystal, no new peak and 
spectral shift were observed, characterizing that both SHG and cTHG emissions are caused by non-resonant (virtual) excitation processes, whose 
emission energies are not related to the resonant band-to-band transition \cite{Shen}. With an increase in the temperature, the SHG intensity decreases faster in comparison to that of cTHG, because phase mismatching for SHG takes place faster than the cTHG. Also in Fig. 2(a), the spectral width is quite broad for the SHG signal, which can be explained by the common nature of harmonic generation that the shorter the wavelength the narrower the bandwidth \cite{Trojnek}. 

Figure 2(b) displays the temperature tuning curve of nonlinear emissions from 20 to 300$^{\circ}$C recorded at the interval of 20$^{\circ}$C without further optical adjustments. As displayed in Fig. 2(b), the integrated SHG (from 600 nm to 750 nm) and cTHG (from 400 nm to 500 nm) intensities show a nearly flat curve at the low-temperature regions. However, at high-temperature regions (from 150 to 300$^{\circ}$C), both the SHG and cTHG intensities drop sharply as temperature increases, and the observed signal intensity is nearly one order of magnitude lower than the signal intensity obtained at room temperature. However, the THG intensity in bulk (pure) diamond crystal gradually decreases as the temperature is increased [Fig. 2(c)]. In the NV diamond, the SHG intensity decreases with an increase in the lattice temperature due to the phase mismatching by the change in the refractive index ($n_{2\omega} \neq n_{\omega}$)  \cite{Boyd}. A dominant source of the refractive index change in semiconductors is the temperature (T) dependent energy bandgap shift [$\Delta E_{g} (T)=E_{g} (T)-E_{g} (0)$] since it acts on the direct and indirect interband electronic transitions, which directly determine the contribution of dielectric function to $n_{2\omega}$ and $n_{\omega}$ even far from the resonance \cite{Yu}. Therefore, using the Bose-Einstein statics and the deformation potential electron-phonon interaction, the temperature dependence of the direct bandgap energy $E_{g}$ can be written as \cite{Giustino}:
\begin{equation} \label{eq2}
\Delta E_{g} (T) = -\alpha[1+2(e^{\hbar \Omega/k_{B} T} - 1)^{-1}].
\end{equation} 
Here, $\alpha$ is a temperature-independent constant, $\hbar \Omega$ is the phonon energy that participates in the electron-phonon renormalization, and $k_{B}$ is the Boltzmann constant.
Based on the good fit result in Fig. 2(b) obtained using Eq. (2), the phonon frequency as a fitting parameter yield is 41.7 THz, which is close to the optical phonon energy in bulk diamond (40 THz) \cite{Solin,Ishioka,Maehrlein} or, more importantly the local vibrational mode related to NV centers (37-40 THz) \cite{Zhang}. On the other hand, the phonon frequency determined from the temperature dependence of the THG signal in bulk (pure) diamond was $\sim$4.0 THz, which would be mainly dominated by the acoustic phonon near the diamond zone center of phonon dispersion \cite{Ulbricht}. The electron-phonon coupling strength for local phonons in the NV diamond is substantially larger than for acoustic phonons, which could explain why long-wavelength optical phonon scattering dominates \cite{Kozk}. Thus, the temperature sensitivity obtained from the SHG of NV diamond ($dI/dT$=0.81\%/$^{\circ}$C) is greater than that from the THG of bulk (pure) diamond ($dI/dT$ = 0.25\%/$^{\circ}$C), showing perspective strategy for developing NLO temperature sensing technologies.

  \begin{figure}[h]
     \includegraphics[width=7.5cm]{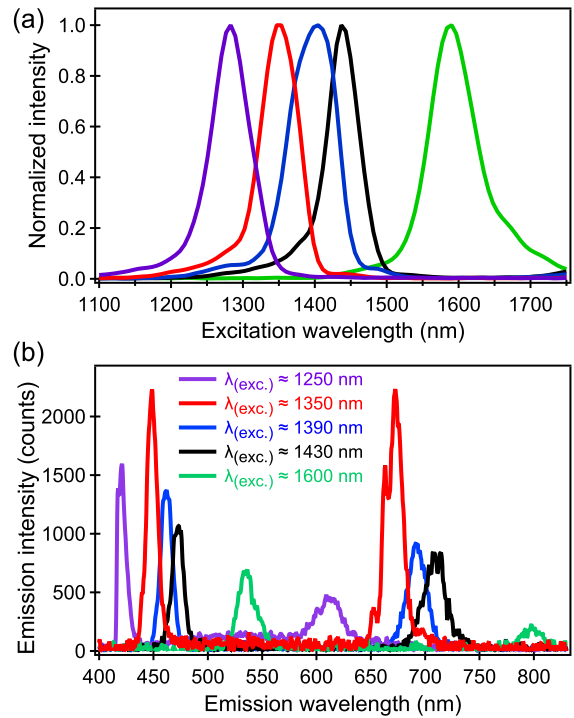}
     \caption{(a) Normalized excitation spectra and (b) nonlinear emission spectra of NV diamond as a function of wavelength.}
     \label{FIG3}
   \end{figure}

In addition to the temperature dependence, we investigated the wavelength tunability of nonlinear emissions from NV diamond. Widely tunable SHG and cTHG have been achieved when the NV diamond crystal is excited by an ultrashort IR laser pulse with an excitation wavelength ranging from 1250 to 1600 nm [Fig. 3(a)]. The corresponding tuning range of SHG was from 625 to 800 nm, and that of cTHG was 415-533 nm as shown in Fig. 3(b). The SHG and cTHG intensity decreases when the excitation IR wavelength is detuned from the initial wavelength centered at $\lambda_{ex}$ $\sim$ 1350 nm possibly due to the reduced electric field intensity and the effect of phase mismatch within NV diamond. Note that fluctuations of intensity appeared in the SHG spectra are possibly due to interference effects from the top and bottom surfaces of a diamond \cite{Kozk}, although it is required to investigate in more details about the wavelength dependent spectral change and temperature sensitivity. Thus, octave-spanning wavelength tuning characteristics of NV diamond crystals offers a promising opportunity for a new type of nonlinear frequency converter based on color centers in diamond crystals, which are helpful to generate a wavelength-tunable diamond laser source.

In conclusion, we investigated the effects of lattice temperature on SHG in NV diamond crystals in the range of 20--300$^{\circ}$. Owing to perfect phase-matching conditions, SHG signals give maximum intensity and the corresponding conversion efficiency was $\eta$ $\sim$ 4.7 $\pm$ 0.9 $\times$ 10$^{-5}$ around room temperature. In contrast, the intensity of both SHG and cTHG decreased sharply at high-temperature regions, which can be explained by the phase mismatch due to temperature-dependent energy band-gap shift. Although the conversion efficiency of SHG and cTHG in NV diamond crystals is one-two orders of magnitude lower than other advanced materials, such as 2D materials \cite{Li2013}, we expect the alignment of NV orientations will enhance the efficiency of NLO effects \cite{Wang2015}. In addition, NV diamond has the advantage that it can be processed into a tip \cite{Pelliccione} to enable a nanometer-scale temperature sensor for various materials. We believe that the present research will provide important evidence for developing diamond-based NLO temperature sensing.


\begin{acknowledgments}
We acknowledge funding support from Core Research for Evolutional Science and Technology program of the Japan Science and Technology (Grant Number: JPMJCR1875).

\end{acknowledgments}


\nocite{*}
\begin{thebibliography}{00}

\bibitem{Benedict}
R. P. Benedict, {\it Fundamentals of Temperature, Pressure, and Flow Measurements} (Wiley, New York, 1969).

\bibitem{Wang2015}
J. Wang, F. Feng, J. Zhang, J. Chen, Z. Zheng, L. Guo, W. Zhang, X. Song, G. Guo, L. Fan, C. Zou, L. Lou, W. Zhu, and G. Wang, Phys. Rev. B \textbf{91}, 155404 (2015).

\bibitem{Pezzagna}
S. Pezzagna, B. Naydenov, F. Jelezko, J. Wrachtrup, and J. Meijer, New J. Phys. \textbf{12}, 065017 (2010).

\bibitem{Breeze}
J. D. Breeze, E. Salvadori, J. Sathian, N. M. Alford, and C. W. M. Kay, Nature \textbf{555}, 493 (2018).

\bibitem{Mildren}
R. P. Mildren and A. Sabella, Opt. Lett. \textbf{34}, 2811 (2009).

\bibitem{OBrien}
J. L. O’Brien, A. Furusawa, and J. Vu\^{c}kovi\'{c}, Nat. Photon. \textbf{3}, 687 (2009).

\bibitem{Sekiguchi}
Y. Sekiguchi, N. Niikura, R. Kuroiwa, H. Kano, and H. Kosaka, Nat. Photon. \textbf{11}, 309 (2017).

\bibitem{Janitz}
E. Janitz, M. K. Bhaskar, and L. Childress, Optica \textbf{7}, 1232 (2020).

\bibitem{Babinec}
T. M. Babinec, B. J. M. Hausmann, M. Khan, Y. Zhang, J. R. Maze, P. R. Hemmer, and M. Lon\^{c}ar, Nat. Nanotechnol.\textbf{5}, 195 (2010).

\bibitem{Mizuochi}
N. Mizuochi, T. Makino, H. Kato, D. Takeuchi, M. Ogura, H. Okushi, M. Nothaft, P. Neumann, A. Gali, F. Jelezko, J. Wrachtrup, and S. Yamasaki, Nat. Photon. \textbf{6}, 299 (2012).

\bibitem{Aharonovich}
 I. Aharonovich, D. Englund, and M. Toth, Nat. Photon. \textbf{10}, 631 (2016).
 
\bibitem{Kim2019}
D. Kim, M. I. Ibrahim, C. Foy, M. E. Trusheim, R. Han, and D. R. Englund. Nat. Electron. \textbf{2}, 284 (2019).

\bibitem{Pelliccione}
 M. Pelliccione, A. Jenkins, P. Ovartchaiyapong, C. Reetz, E. Emmanouilidou, N. Ni, and A. C. B. Jayich, Nat. Nanotechnol. \textbf{11}, 700 (2016).
 
\bibitem{Dolde}
F. Dolde, H. Fedder, M. W. Doherty, T. N\"{o}bauer, F. Rempp, G. Balasubramanian, T. Wolf, F. Reinhard, L. C. L. Hollenberg, F. Jelezko, and J. Wrachtrup, Nat. Phys. \textbf{7}, 459 (2011).

\bibitem{Maze2008}
J. R. Maze, P. L. Stanwix, J. S. Hodges, S. Hong, J. M. Taylor, P. Cappellaro, L. Jiang, M. V. G. Dutt, E. Togan, A. S. Zibrov, A. Yacoby, R. L. Walsworth, and M. D. Lukin, Nature \textbf{455}, 644 (2008).

\bibitem{Kucsko}
G. Kucsko, P. C. Maurer, N. Y. Yao, M. Kubo, H. J. Noh, P. K. Lo, H. Park, and M. D. Lukin, Nature \textbf{500}, 54 (2013).

\bibitem{Clevenson}
H. Clevenson, M. E. Trusheim, C. Teale, T. Schr\"{o}der, D. Braje, and D. Englund, Nat. Phys. \textbf{11}, 393 (2015).

\bibitem{Trojnek}
F. Troj\'{a}nek, K. Z\'{i}dek, B. Dzurn\'{a}k, M. Koz\'{a}k, and P. Mal\'{y}, Opt. Exp. \textbf{18}, 1349 (2010).

\bibitem{Abulikemu}
A. Abulikemu, Y. Kainuma, T.  An, and M. Hase, ACS Photon. \textbf{8}, 988 (2021).

\bibitem{Motojima}
M. Motojima, T. Suzuki, H. Shigekawa, Y. Kainuma, T. An, and M. Hase, Opt. Exp. \textbf{27}, 32217 (2019).

\bibitem{Angerer}
A. Angerer, K. Streltsov, T. Astner, S. Putz, H. Sumiya, S. Onoda, J. Isoya, W. J. Munro, K. Nemoto, J. Schmiedmayer, and J. Majer, Nat. Phys. \textbf{14}, 1168 (2018).

\bibitem{Shen}
Y. R. Shen, {\it The Principles of Nonlinear Optics} (Wiley, New York, 1984), Chap. 7.

\bibitem{Prananto}
D. Prananto, Y. Kainuma, K. Hayashi, N. Mizuochi, K. Uchida, and T. An, Phys. Rev. Applied \textbf{16}, 064058 (2021).

\bibitem{Yurov}
V. Yu. Yurov, E. V. Bushuev, A. F. Popovich, A. P. Bolshakov, E. E. Ashkinazi, and V. G. Ralchenko, J. Appl. Phys. \textbf{122}, 243106 (2017).

\bibitem{Boyd}
R. W. Boyd, {\it Nonlinear Optics 3rd ed.} (Academic Press, 2008) , Chap. 2.

\bibitem{Yu}
P. Yu and M. Cardona, {\it Fundamentals of Semiconductors} (Springer, 2005) , Chap. 6.

\bibitem{Giustino}
F. Giustino, S. G. Louie, and M. L. Cohen, Phys. Rev. Lett. \textbf{105}, 265501 (2010).

\bibitem{Solin}
S. A. Solin, and A. K. Ramdas, Phys. Rev. B \textbf{1}, 1687 (1970).

\bibitem{Ishioka}
K. Ishioka, M. Hase, M. Kitajima, and H. Petek, Appl. Phys. Lett. \textbf{89}, 231916 (2006).

\bibitem{Maehrlein}
S. Maehrlein, A. Paarmann, M. Wolf, and T. Kampfrath, Phys. Rev. Lett. \textbf{119}, 127402 (2017). 

\bibitem{Zhang}
J. Zhang, C.-Z. Wang, Z. Z. Zhu, and V. V. Dobrovitski, Phys. Rev. B \textbf{84}, 035211 (2011).

\bibitem{Ulbricht}
R. Ulbricht, S. Dong, I-Ya Chang, B. M. K. Mariserla, K. M. Dani, K. Hyeon-Deuk, and Z.-H. Loh, Nat. Commun. \textbf{7}, 13510 (2016).

\bibitem{Kozk}
M. Koz\'{a}k, F. Troj\'{a}nek, B. Rezek, A. Kromka, and P. Mal\'{y}, Physica E \textbf{44}, 1300 (2012).

\bibitem{Li2013}
Y. L. Li, Y. Rao, K. F. Mak, Y. M. You, S. Y. Wang, C. R. Dean, and T. F. Heinz, Nano Lett. \textbf{13}, 3329 (2013).

\end{thebibliography}
\end{document}